\documentclass[10pt,epsf]{article}

\usepackage{graphicx}
\usepackage{indentfirst}
\usepackage{color}
\usepackage{subfigure}
\usepackage{caption}

\begin{document}

\title{\bf{Instability of Rotating Anti-de Sitter Black Holes}}

\date{}
\maketitle

\begin{center}\author{Bogeun Gwak}$^a$\footnote{rasenis@sogang.ac.kr} and \author{Bum-Hoon Lee}$^a$$^b$\footnote{bhl@sogang.ac.kr}\\
\vskip 0.25in
$^{a}$\it{Center for Quantum Spacetime, Sogang University, Seoul 121-742, Korea}\\
$^{b}$\it{Department of Physics, Sogang University, Seoul 121-742, Korea} \end{center} \vskip 0.6in

{\abstract
{We study the instability of higher-dimensional rotating anti-de Sitter black holes through fragmentation. Fragmentation occurs when black holes rotate too fast to sustain their horizon, and then the black holes are broken into small pieces. Using this process, we test the stability of AdS black holes and discover the dynamical upper bounds of the angular momentum and the cosmological constant. We show that AdS black holes can exist stably within limited parameter ranges in the general dimensions. The parameter ranges are obtained numerically in terms of angular momentum and cosmological constant.}}\\

\thispagestyle{empty}
\newpage
\setcounter{page}{1}
\section{Introduction}
A Kerr black hole is a rotating black hole as a solution to Einstein gravity. The angular momentum from the rotation is bounded, and the bound prevents a naked singularity (NS) in 4-dimensional spacetime. We call this bound the Kerr bound. In higher-dimensional spacetime, a Kerr black hole generalizes into a Myers-Perry\,(MP) black hole\,\cite{Myers:1986un}\,. An MP black hole has an outer horizon for any angular velocity value in 6 dimensions and higher\,\cite{Emparan:2008eg}\,. An MP black hole with an infinitely large angular velocity is called an ultra-spinning black hole, which is hard to achieve physically with particle absorption\,\cite{Gwak:2011rp}\,. The instability of an MP black hole increases as the angular velocity increases. The growing instability acts as the natural upper bound of the angular momentum\,\cite{Emparan:2007wm}\,. Since the bound comes from instability or black hole motion, it is called a dynamic Kerr bound. In the law of thermodynamics, black hole entropy is proportional to the area of the black hole\,\cite{Bekenstein:1973ur}\,. The area of the black hole decreases as the angular momentum increases. For a given angular momentum and mass, the black hole phases remain in a large entropy state. An ultra-spinning black hole is unstable because of its small entropy and undergoes a phase transition\,\cite{Emparan:2007wm,Dias:2009iu} or fragmentation\,\cite{Emparan:2003sy}\,. Fragmentation is a phase transition in which an MP black hole is broken into many black holes because the entropy of the initial black hole is smaller than the sum of the entropies of the final fragmented black holes. An MP black hole is unstable in large angular velocity, and the upper bound of the angular velocity from fragmentation exists.

Black holes can extend to include a cosmological constant. A black hole with a negative cosmological constant is asymptotic anti-de Sitter (AdS) spacetime. The physics of an AdS black hole can be interpreted in terms of conformal field theory\,(CFT) using AdS/CFT duality. AdS/CFT duality can describe black hole entropy with CFT defined on the spacetime boundary\,\cite{Maldacena:1997re}\,. Black hole entropy was derived with the logarithm for counting D-brane microstates\,\cite{entD}\,. The gravity/gauge duality progresses to discover the duality of asymptotically flat spacetime, such as the Kerr/CFT correspondence\,\cite{kerrCFT} and the RN/CFT correspondence\,\cite{Hartman:2008pb}. From the gauge/gravity duality, changes in spacetime, such as a black hole phase transition, can appear as CFT defined on the spacetime boundary. CFT that lives on the boundary should change when a black hole undergoes instability or the phase changes\,\cite{Gwak:2012hq}\,. Investigation of AdS black hole instability shows the relationship between the AdS black-hole phase and CFT on the boundary, which can be elucidated using the well-defined AdS/CFT duality.

A rotating AdS black hole\,\cite{Hawking:1998kw,Carter:1968ks}\, is unstable due to superradiance\,\cite{Cardoso:2004hs,Cardoso:2006wa,Cardoso:2004nk} and perturbation\,\cite{Kunduri:2006qa,Kodama:2009rq,Monteiro:2009ke}\,. Superradiance instability occurs in $\Omega_H \ell>1\,$ where $\ell$\, and $\Omega_H\,$ are the AdS radius and the angular velocity of the AdS black hole, respectively~\cite{Cardoso:2004hs, Cardoso:2006wa, Kunduri:2006qa}\,. A rotating AdS black hole extracts its rotating energy from gravitational wave scattering. The gravitational wave should co-rotate with the black hole and should satisfy $\omega < n \Omega_H$\, where $n\,$ is the angular momentum quantum number of the wave. After the rotating energy is extracted, the gravitational wave is reflected in the AdS boundary. The reflected wave extracts the rotation energy from the black hole, and the black hole becomes unstable when $\Omega_H \ell >1$\,. CFT is defined in the AdS boundary of the black hole, and it appears in a rotating Einstein universe\,\cite{Hawking:1998kw, Hawking:1999dp}\,. There is a BPS-type upper bound in the angular momentum in a single spinning AdS black hole\,\cite{Hawking:1998kw,Chrusciel:2006zs}\,, but there is no upper bound in an asymptotically flat MP black hole. A single spinning MP black hole has ultra-spinning instability in a large angular momentum. Instability also occurs in a higher-dimensional rotating AdS black hole, as an extension of a 4-dimensional rotating AdS black hole, and a non-rotating AdS black hole~\cite{Monteiro:2009ke,Prestidge:1999uq,Dias:2010gk}.

In this paper, we investigate the non-perturbed instability of higher-dimensional rotating AdS black holes through fragmentation. Fragmentation instability\,\cite{Emparan:2003sy}\, generalizes into an asymptotic AdS black hole and acts as the upper bounds of black hole parameters. Called a dynamic Kerr bound, the upper bounds show that an AdS black hole can exist stably within limited parameter ranges. The fragmentation instability will be presented in the phase diagram including superradiance instability. We find that the unstable region exists in stable region under superradiance. The minimum parameter ranges of dynamic Kerr bounds were obtained numerically. Mass loss and multiple fragmentation are considered from a general point of view.

This paper is organized as follows. In section~2, we provide a review of higher-dimensional rotating AdS black holes. A black hole has two angular velocity upper bounds related to its mass and cosmological constant. In section~3, we compute the unstable parameter ranges for fixed mass and cosmological constant slices with fragmentation in general dimensions. In fragmentation, we consider the mass loss from the gravitational radiation and multiple fragmentation. In section~4, we conclude and discuss our results.

\section{Rotating Anti-de Sitter Black Holes}

The rotating AdS black hole with a single rotation in Anti-de Sitter spacetime\,\cite{Hawking:1998kw,Carter:1968ks} is
\begin{eqnarray}
\label{eq:metric}
&&ds^2 = -\frac{\Delta_r}{\rho^2}\left(dt-\frac{a \sin^2\theta}{\Xi}d\phi\right)^2+\frac{\sin^2\theta \Delta_\theta}{\rho^2}\left(\frac{r^2+a^2}{\Xi}d\phi -a dt\right)^2 +\frac{\rho^2}{\Delta_r}dr^2+\frac{\rho^2}{\Delta_\theta}d\theta^2 +r^2\cos^2\theta d\Omega^2_{D-4}\,,\nonumber\\
&&\Delta_r = \left({r^2+a^2}\right)\left({1+\frac{r^2}{\ell^2}}\right)-\frac{\mu}{r^{D-5}}\,,\,\,\Delta_\theta = 1-\frac{a^2}{\ell^2}\cos^2\theta\,,\,\, \rho^2 = r^2 + a^2\cos^2\theta\,,\,\, \Xi=1-\frac{a^2}{\ell^2}\,,
\end{eqnarray} 
where the AdS curvature radius is $\ell$\,, the mass parameter is $\mu\,$, and the spin parameter\,(spin) is $a\,$. The metric is the solution of the Einstein equation with a negative cosmological constant, $R_{\mu\nu}=-(D-1)\ell^{-2}g_{\mu\nu}\,.$ The black hole mass $M\,$ and the angular momentum $J$ are
\begin{eqnarray}
\label{eq:Kmass}
M=\frac{A_{D-2}}{8\pi G \Xi^2}\left(1+\frac{D-4}{2}\Xi \right)\mu\,,\,\,
J=\frac{A_{D-2}}{8\pi G \Xi^2}\mu a\,,\,\,
A_{D-2}=\frac{2\pi^{(D-2)/2}}{\Gamma((D-1)/2)}\,,
\end{eqnarray}
where $A_{D-2}$ is the volume of a unit radius (D-2)-sphere, and $G$ is Newton's constant. The metric has an upper bound for the BPS-like angular momentum bound\,\cite{Hawking:1998kw,Chrusciel:2006zs} from eq.~(\ref{eq:metric})
\begin{eqnarray}
|a| < \ell\,.
\end{eqnarray}
In the limit $|a|\rightarrow \ell$, the horizon is divergent and vanished, and $\Xi\rightarrow 0\,$. Then, the solution becomes a naked singularity. The angular velocity with respect to a non-rotating frame at infinity is\,\cite{Gibbons:2004ai,Caldarelli:1999xj}
\begin{eqnarray}
\Omega_H=\frac{a}{r_h^2+a^2}\left(1+\frac{r_h^2}{\ell^2}\right)\,.
\end{eqnarray}
The superradiance instability occurs over $\Omega_H\,\ell>1$\,\cite{Dias:2009iu,Cardoso:2004hs,Kunduri:2006qa,Kodama:2009rq,Dias:2010gk}\,.
The black hole entropy is proportional to the horizon area $\mathcal{A}$\,\cite{Gibbons:2004ai}
\begin{eqnarray}
\mathcal{A} = \frac{A_{D-2}(r_h^2+a^2)r_h^{D-4}}{\Xi}\,,
\end{eqnarray}
which depends on the AdS radius $\ell\,$ in $\Xi\,$.

\section{Instability from Fragmentation}
The angular momentum affects black hole instability. With a large angular momentum, a black hole needs too large a centrifugal force to maintain the horizon, so the black hole breaks apart into two identical black holes, carrying away the spin as the orbital angular momentum\,\cite{Emparan:2003sy}\,. The initial state is a black hole that has mass $M$ and angular momentum $J$\,. Growing instability in a high spin breaks the initial black hole into the black holes of the final state. We define the final state in which there are two identical black holes with mass $m$ and linear momentum $\pm J/2R$ with the impact parameter $R$ from the center of the mass. If we assume the final black holes as a particle moving in AdS Schwarzchild black hole spacetime, the particle energy $E$ and angular momentum $L$ are related as
\begin{eqnarray}
\label{eq:energy10}
E^2=\left(1+\frac{R^2}{\ell^2}-\frac{\mu}{R^{D-3}}\right)\left(m^2+\frac{L^2}{R^2}\right).
\end{eqnarray} 
A reasonable value of the impact parameter $R$ is at least smaller than the horizon radius, which is from the parallel two-dimensional area $A_{\|}$ at the point of view in the transverse sphere $\Omega_{d-4}\,$. Thus, the impact parameter is smaller than the horizon radius~\cite{Emparan:2003sy}\,,
\begin{eqnarray}
\label{eq:equality3}
R\leq \sqrt{\frac{r_h^2+a^2}{\Xi}}\,.
\end{eqnarray}
If we choose equality in eq.~(\ref{eq:equality3})\,, the impact parameter is expanded in large AdS radius
\begin{eqnarray}
R^2\approx(r_h^2+a^2)+\frac{a^2(r_h^2+a^2)}{\ell^2}\,.
\end{eqnarray}
We suppose that a small cosmological constant
\begin{eqnarray}
\ell \gg R\,,
\end{eqnarray}
where AdS spacetime effect in eq.~(\ref{eq:energy10}) can be approximated to the case of asymptotically flat spacetime~\cite{Emparan:2003sy}\,. After fragmentation, the final state black holes are infinitely far from each other. Thus, interactions between black holes can be ignored, and we can treat them as two independent single black holes. The mass of the initial state is conserved to the mass and linear momentum of the final state,
\begin{eqnarray}
M=2\sqrt{m^2+\frac{J^2}{4R^2}}\,,
\end{eqnarray}
in which the mass of final state is obtained
\begin{eqnarray}
\label{eq:mass1}
m=\frac{1}{2}\sqrt{M^2-\frac{J^2}{R^2}}\,.
\end{eqnarray}
Applying eq.~(\ref{eq:Kmass}) into eq.~(\ref{eq:mass1})\,, the mass is rewritten in terms of mass parameters 
\begin{eqnarray}
\frac{A_{D-2}}{8\pi G}\left(1+\frac{D-4}{2}\right)\mu'=\frac{1}{2}\sqrt{\left(\frac{A_{D-2}}{8\pi G \Xi^2}\left(1+\frac{D-4}{2}\Xi \right)\mu\right)^2-\frac{1}{R^2}\left(\frac{A_{D-2}}{8\pi G \Xi^2}\mu a\right)^2}\,,
\end{eqnarray}
where the mass parameter and the spin of the initial black hole are $\mu$ and $a$\,, and the mass parameter of the final black holes is $\mu'$\,. Note that the spin of the final black hole is zero and is transferred to the linear momentum of the final black holes. The mass parameter is obtained such that
\begin{eqnarray}
\mu'=\frac{\mu}{(D-2)\Xi^2}\sqrt{\left(1+\frac{D-4}{2}\Xi \right)^2-\frac{a^2}{R^2}}\,.
\end{eqnarray}
The area of the final state is proportional to the mass parameter $\mu'\,$. 
The equality in eq.~(\ref{eq:equality3}) maximizes the entropy of the final state black holes, whose masses are
\begin{eqnarray}
\label{eq:massp1}
&&\mu'=\frac{\mu}{(D-2)\Xi^2}\sqrt{\left(1+\frac{D-4}{2}\Xi \right)^2-\frac{\Xi\,a^2}{r_h^2+a^2}}\,,
\end{eqnarray} 
where the final black hole mass parameter $\mu'$ is written in terms of initial state black hole parameters. According to the second law of thermodynamics, a larger entropy state is stable and preferred. We can assume that black holes also tend to move to a larger entropy state; thus, fragmentation occurs when the entropy in the final state is larger than that in the initial state. The entropy of a black hole is proportional to the black hole horizon area, which is invariant under boost~\cite{Horowitz:1997fr}. The ratio of the initial and final states is
\begin{eqnarray}
\label{eq:entratio3}
\frac{\mathcal{A}_f}{\mathcal{A}_i}=\frac{2r'^{D-2}_h \Xi}{(r^2_h+a^2)r^{D-4}_h}\,,
\end{eqnarray}
where the black hole is fragmented into two black holes when the ratio is larger than 1, and we suppose the black hole is unstable and define the angular momentum as a dynamical Kerr bound. Dynamical Kerr bounds are difficult to solve; therefore, we compute the points using a numerical method. The dimensionality can be classified into three cases: $D=4\,$, $D=5$\,, and $D \geq 6\,$.

\subsection{Analytical Approximation in General Dimensions}
Before approaching black holes with the numerical method, we test the stability of black holes within the large and small parameter limits of the general dimensions. Three parameters describe black holes, so stability will be discussed in the $\ell$ and $\mu$ fixed slices. First, a non-rotating small mass black hole is an AdS Schwarzschild black hole in the $\mu \ll 1$ and $a \ll 1$ limits. In our approach, the stability of the black hole depends on the entropy ratio. Since black hole entropy can be obtained from the horizon, we need to find the horizon. The horizon $r_h$ is a solution that
\begin{eqnarray}
\label{eq:hrzeq1}
\frac{r_h^{D-1}}{\ell^2}+(1+\frac{a^2}{\ell^2})r_h^{D-3}+a^2 r_h^{D-5}-\mu=0\,,
\end{eqnarray}
which is from eq.~(\ref{eq:metric})\,. The lowest power of the horizon is the most dominant term in the small mass limit, so the horizon is approximately $r_h=(\mu)^{1/{D-3}}$\,. The mass of the black hole in the final state is $\mu'=\mu/2$ from eq.~(\ref{eq:massp1}) in the limit. Under these conditions, the entropy ratio is $\mathcal{A}_f/\mathcal{A}_i=(1/2)^{1/D-3}<1$\,. Therefore, an AdS Schwarzschild black hole with a small mass is stable under fragmentation. In contrast, the highest power of $r_h$ is dominant in large mass $\mu\gg 1$ without rotation. The horizon is approximately $r_h=(\mu \ell^2)^{1/{D-1}}$\,, and then the ratio is $\mathcal{A}_f/\mathcal{A}_i=2^{1/D-1}>1$\,. An AdS Schwarzschild black hole with a large mass is unstable. It is not seen in a flat Schwarzschild black hole. Thus, the instability comes from the negative cosmological constant effect. In addition, we can expect a dynamic Kerr bound in the zero angular momentum.

The angular momentum of a black hole is limited under $a<\ell$. In addition, 4- and 5-dimensional black holes have Kerr bounds in a small angular momentum, so let us check the stability of black holes with small mass and $a\sim\ell$\, in 6 dimensions and more. Note that $\Xi \rightarrow 0$\, in the limit of $a \rightarrow \ell\,$. Similar to the AdS Schwarzschild black hole with small mass, the horizon is given as $(\mu/a^2)^{1/{D-5}}\,$, but the entropy behaves differently from the zero momentum. The entropy ratio is rewritten from eq.~(\ref{eq:entratio3})
\begin{eqnarray}
\label{eq:entratio2}
\frac{\mathcal{A}_f}{\mathcal{A}_i}=\frac{2(\mu'/\ell^2)^{D-2} \Xi}{((\mu/\ell^2)^2+\ell^2)(\mu/\ell^2)^{D-4}}\,,
\end{eqnarray}
where the numerator is finite for given parameter, so the entropy behavior is decided by $\Xi$. However, the final black hole mass $\mu'$ has a $\Xi$ contribution in eq.~(\ref{eq:massp1})
\begin{eqnarray}
\mu'=\frac{\mu}{(D-2)\Xi^2}\sim\Xi^{-2}\,.
\end{eqnarray}
Substituted in the final mass of the black hole, the entropy ratio depends completely on $\mathcal{A}_f / \mathcal{A}_i \sim \Xi^{(4-D)/(D-5)} \gg 1\,$. Thus, the black hole becomes unstable before $a\sim \ell$ BPS-like bound. The centrifugal force acts too strongly on the black hole for it to maintain its horizon. As a result, the black hole fragments, and we also expect dynamic Kerr bounds between $a=0$ and $a\sim\ell$\,. Detailed behaviors of dynamic Kerr bounds are discovered with numerical analysis in the following sections.

Now, we redefine the dimensionless coordinates for the numerical analysis as
\begin{eqnarray}
\tilde{t}=\frac{t}{\mu^{1/D-3}}\,,\,\,\,\,\tilde{r}=\frac{r}{\mu^{1/D-3}}\,,\,\,\,\,\tilde{a}=\frac{a}{\mu^{1/D-3}}\,,\,\,\,\,\tilde{L}=\frac{\ell}{\mu^{1/D-3}}\,,
\end{eqnarray}
where $\tilde{\mu}=1$\,. The mass parameter of the final black hole is rewritten from eq.~(\ref{eq:massp1})\,,
\begin{eqnarray}
\label{eq:massp3}
&&\tilde{\mu}'=\frac{1}{(D-2)\tilde{\Xi}^2}\sqrt{\left(1+\frac{D-4}{2}\tilde{\Xi} \right)^2-\frac{\tilde{\Xi}\,\tilde{a}^2}{\tilde{r}_h^2+\tilde{a}^2}}\,,
\end{eqnarray} 
The ratio of the initial and final states is also rewritten from eq.~(\ref{eq:entratio3})\,,
\begin{eqnarray}
\label{eq:entratio5}
\frac{\mathcal{A}_f}{\mathcal{A}_i}=\frac{2\tilde{r}'^{D-2}_h \tilde{\Xi}}{(\tilde{r}^2_h+\tilde{a}^2)\tilde{r}^{D-4}_h}\,,
\end{eqnarray}
Therefore, phase diagrams for AdS black holes for $\tilde{\mu}=1$ is represented in terms of $(1/\tilde{\ell},\tilde{a})$ parameters.

\subsection{The 4-dimensional AdS Rotating Black hole}
The 4-dimensional rotating AdS black hole has two bounds: One comes from the Kerr bound to prevent naked singularity. This bound is seen only in 4 and 5 dimensions.
\begin{figure}[h]
\centering{\includegraphics[scale=0.6,keepaspectratio]{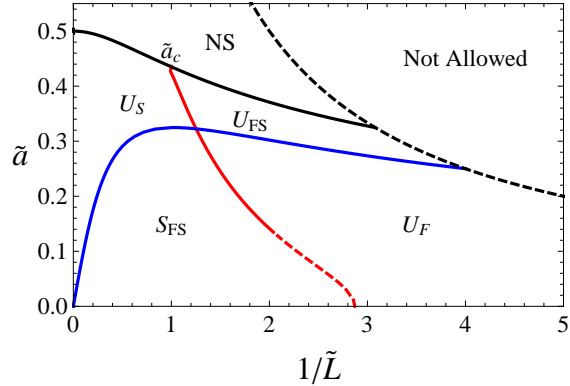}
\centering\caption{{\small The phase diagrams for $D=4$. The thick black line represents the Kerr bound. The thick blue line is the superradiance instability, and the dashed black line is the BPS-like $a=|\ell|\,$ bound. The thick red line means unstable from the fragmentation. The dashed red line means unstable from the fragmentation for the large black hole, so it is outside of our approximation, and we do not consider in this region. The black hole in the region $S_{FS}$ is stable under both fragmentation and superradiance. The black hole in the region $U_{F}$ is unstable under fragmentation. The black hole in the region $U_{S}$ is unstable under superradiance. The black hole in the region $U_{FS}$ is unstable under both fragmentation and superradiance. The shapes and colors of lines are the same in our paper.}}
\label{fig1}}
\end{figure}
The other bound is a BPS-like bound from the cosmological constant, $a=|\ell|\,$. The BPS-like bound exists in any spacetime dimension, so in the 4-dimensional rotating AdS black hole, the spin parameter has these two boundaries. Since the black hole spin parameter is limited under the Kerr bound, we should discuss stability below the Kerr bound. 
The detailed phase structure is given in Figure~\ref{fig1}\,. The dynamic Kerr bound is represented on the $\tilde{a}-1/\tilde{L}$ diagram by red solid line. The black hole is stable in $S_{FS}$ region and unstable in $U_F$, $U_S$, $U_{FS}$\,. The regions $U_F$ is stable under superradiance, but the black hole should be stable under fragmentation, so the stable region of the black hole becomes smaller to the region $S_{FS}$. With the limit of $1/\ell\rightarrow 0$, the black hole behaves like a Kerr black hole and is stable for arbitrary values of $a$ under fragmentation. The fragmentation instability starts at the critical value of $\tilde{a}_c$\,. In non-zero $1/\ell$\,, the AdS effect increases the centrifugal force of the black hole and gives it an unstable spin parameter $\tilde{a}_c$\,; thus, a dynamic Kerr bound appears. Our approximation is not valid in the dashed red line, we remain it for overall tendency.

\subsection{The AdS Rotating Black Hole in 5 Dimensions and over}
\begin{figure}[h] \subfigure[The diagram for $D=5$\,.]{\includegraphics[scale=0.6,keepaspectratio]{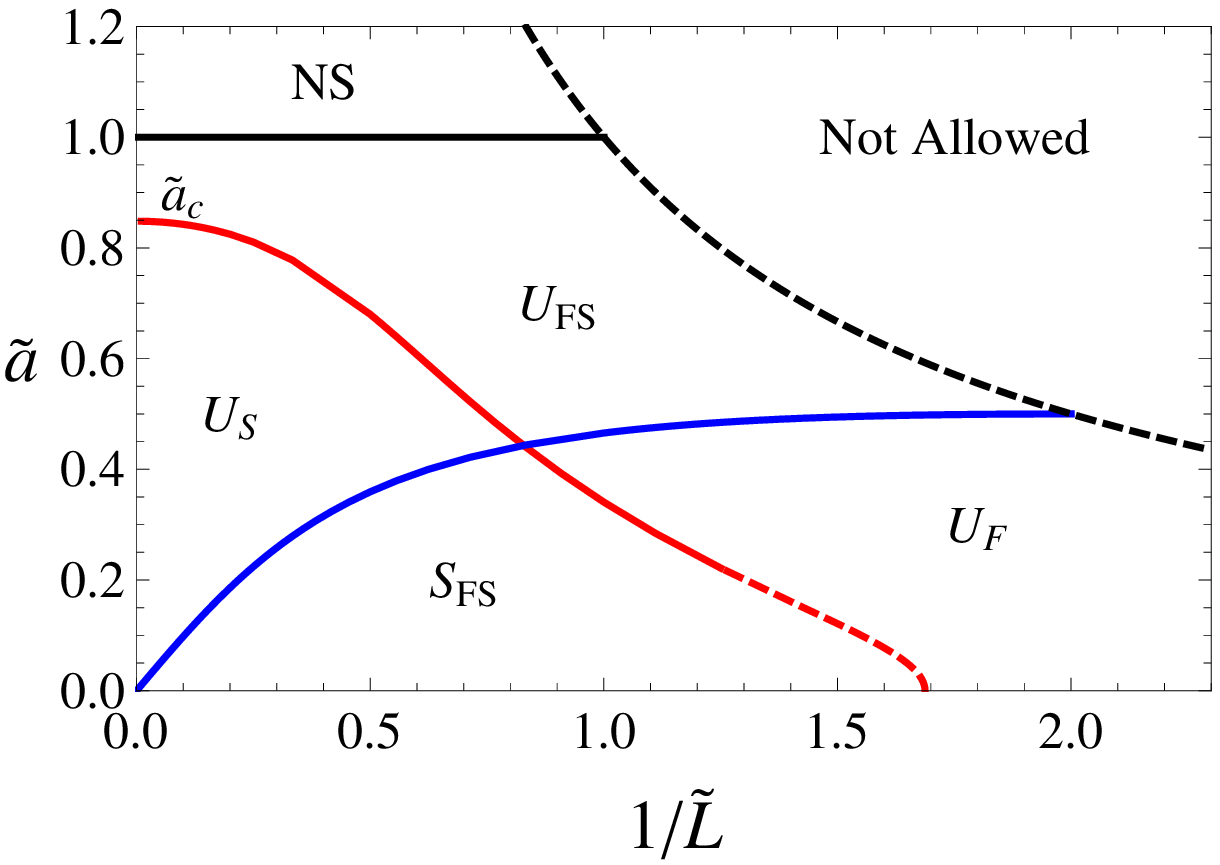}} \hfill \subfigure[The diagram for $D=6$\,.]{\includegraphics[scale=0.6,keepaspectratio]{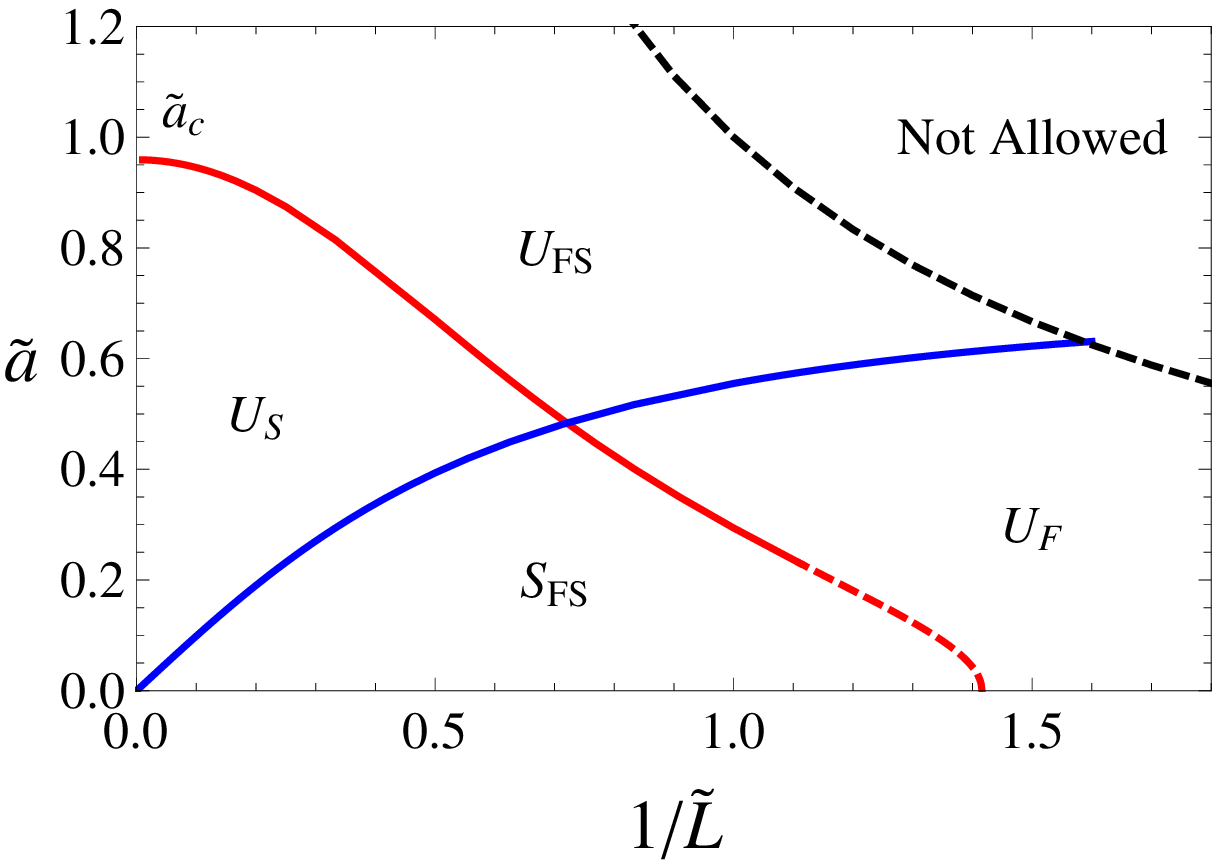}} \caption{{\small The phase diagrams for $D=5$ and $D=6$. The phase diagrams over 6-dimensional cases are similar to 6-dimensional diagram. Once again, the dashed red lines are not valid in our approximation.}} \label{fig2} \end{figure}

The instability of a rotating black hole in 5 Dimensions and over is different from that of the 4-dimensional case. The 5-dimensional AdS black hole still has an extremality like 4-dimensional case. The size of an extremal black hole is zero. From the minimum, the spin parameter is bounded at
\begin{eqnarray}
a\leq\sqrt{\mu}\,,
\end{eqnarray}
where the condition is independent of the cosmological constant, so the maximum values of the spin parameter appear as a straight line in the diagram. The phase structure is shown in Figure~\ref{fig2}-(a)\,. Although the power of gravity becomes higher than that for a 4-dimensional black hole, black holes are fragmented in a sufficient angular momentum even if zero cosmological constant. The fragmentation instability starts at the spin parameter $\tilde{a}_c$\,. The fragmentation gives upper bound for $\tilde{a}$ for given a cosmological constant below the Kerr bound and acts as a dynamic upper bound. Different from the 4-dimensional case, in 5 dimensions a dynamic Kerr bound appears before the Kerr bound and BPS-like bounds. In more than 4 dimensions, the centrifugal force affects the instability of black holes more than before, because the critical value $a_c$ appears below the Kerr and BPS-like bounds.
The AdS black hole is stable in the region $S_{FS}$\, and unstable in the region $U_F$, $U_S$, and $U_{FS}$\,. 

In more than 5 dimensions, a rotating AdS black hole has an outer horizon for any value of the spin parameter $\tilde{a}\,$. The horizon is a solution to eq.~(\ref{eq:hrzeq1})\,. Because there is a proper $r_h$ for any values of $a$\,, the equation always has a solution. As a result, the black hole has no Kerr bounds, and the dynamic Kerr bound competes only with the BPS-like bound. The instability behavior is similar to that of 5-dimensional cases as shown in the diagram in Figure~\ref{fig2}-(b)\,.
On the limit of $1/\tilde{L} \rightarrow 0$, the instability occurs at a ratio of $a_c/r_h$\,\cite{Emparan:2003sy}\,,
\begin{eqnarray}
\frac{a_c}{r_h}\geq 1.36\,\,\textrm{in}\,\,D=6\,,\,\,\frac{a_c}{r_h}\geq 1.26\,\,\textrm{in}\,\,D=7\,,\,\,\frac{a_c}{r_h}\geq 1.20\,\,\textrm{in}\,\,D=8\,.
\end{eqnarray}
Therefore, the flat case limits correspond to the cases of MP black hole. These dynamic Kerr bounds are extended to negative cosmological cases as shown in red lines of Figure~\ref{fig2}-(b)\,. In the AdS spacetime, the centrifugal force makes the black hole more unstable, because the dynamic Kerr line goes down and appears in smaller values of the spin parameters.

\subsection{Small Mass Loss in Fragmentation}

The black hole can lose its small mass in the fragmentation. We define the infinitesimal mass $\delta M$\,. The mass will be radiated as the gravitational wave. The entropy of the wave is treated as a non-rotating black hole. The entropy of the black hole is the maximum for the given mass, so the effects of the mass loss can be obtained. The total mass becomes smaller than before.
\begin{eqnarray}
M-\delta M=2\sqrt{m^2+\frac{J^2}{4R^2}}\,,
\end{eqnarray}
where the energy radiated is approximately given as $\delta \mu$\,. The mass and angular momentum are redefined as the mass and spin parameters. The non-rotating black hole mass $\mu'$ is in its final state.
\begin{eqnarray}
\mu'\simeq\frac{\mu}{(D-2)\Xi^2}\left[\sqrt{\left(1+\frac{D-4}{2}\Xi \right)^2-\frac{a^2}{R^2}}-\frac{\left(1+\frac{D-4}{2}\Xi \right)^2}{\sqrt{\left(1+\frac{D-4}{2}\Xi \right)^2-\frac{a^2}{R^2}}}\frac{\delta \mu}{\mu}\right]\,,
\end{eqnarray}
where the infinitesimal mass change in the final state is
\begin{eqnarray}
\delta \mu'=\frac{\left(1+\frac{D-4}{2}\Xi \right)^2}{(D-2)\Xi^2\sqrt{\left(1+\frac{D-4}{2}\Xi \right)^2-\frac{\Xi a^2}{r_h^2+a^2}}}\,\delta\mu>0\,,
\end{eqnarray}
where the impact parameter is inserted. The horizon change on the final state is,
\begin{eqnarray}
\delta r_h'=-\frac{r_h^4\ell^2}{(D-1)r_h^{D+2}+(D-3)r_h^D\ell^2}\delta \,\mu'<0\,.
\end{eqnarray}
The entropy ratio between the initial and final states is rewritten as
\begin{eqnarray}
\frac{\mathcal{A}_f}{\mathcal{A}_i}=\frac{\mathcal{A}_{2BH}+\mathcal{A}_{\delta M}}{\mathcal{A}_i}\simeq\frac{2r'^{D-2}_h}{(r^2_h+a^2)r^{D-4}_h}+\frac{2(D-2)r'^{D-3}_h\,\delta r_h'+\mathcal{A}_{\delta M}}{(r^2_h+a^2)r^{D-4}_h}\,.
\end{eqnarray}
The gravitational wave entropy is $\mathcal{A}_{\delta M}$\,, and the decrease of entropy treated as the non-rotating black hole mass $\delta M$ is related to $\delta r'_h$\,. The gravitational wave entropy is smaller than that of the non-rotating black hole in the same mass $\delta M$, so the total entropy of the final state decreases in comparison with the no-mass loss case. This means that
\begin{eqnarray}
2(D-2)r'^{D-3}_h\,\delta r_h'+\mathcal{A}_{\delta M}<0\,.
\end{eqnarray}
The phase transition should be retarded by gravitational radiation. The radiation makes that the fragmentation occurs larger spin parameter, and the unstable red boundary will move upside in the Figure~\ref{fig1}, \ref{fig2}\,. This result is the same as the fragmentation of static black holes\cite{Ahn:2014fwa}\,.

\subsection{Fragmentation to More than Two Black Holes}
\begin{figure}[h] \subfigure[The diagram for $D=4$.]{\includegraphics[scale=0.6,keepaspectratio]{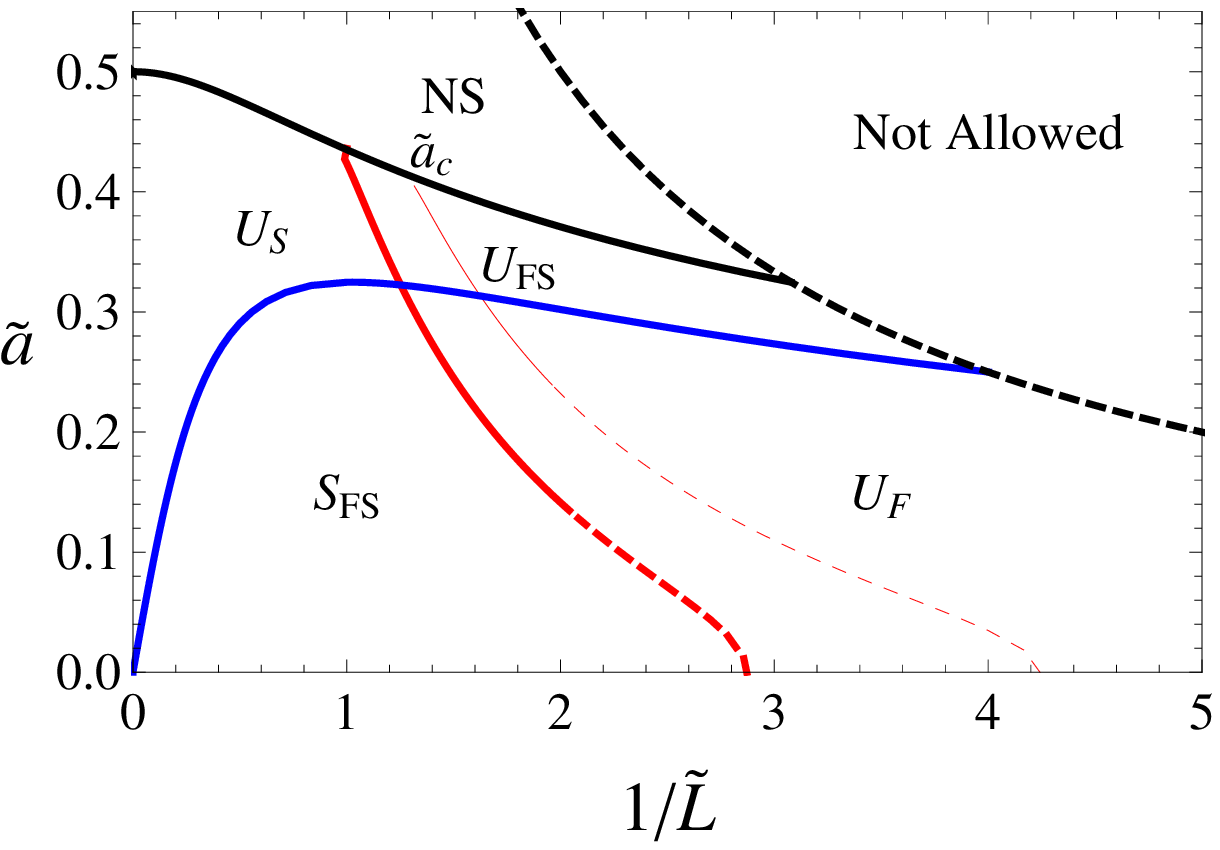}} \hfill \subfigure[The diagram for $D=6$.]{\includegraphics[scale=0.6,keepaspectratio]{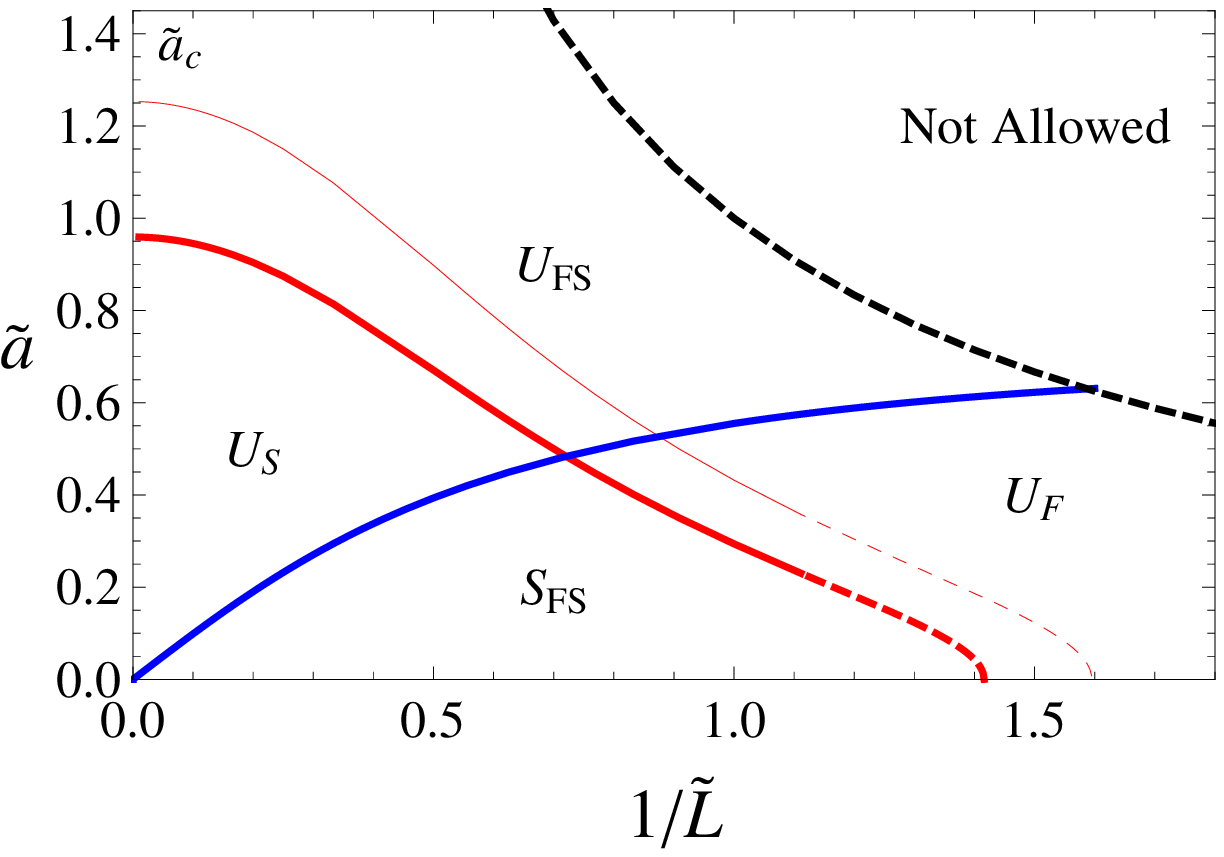}} \caption{{\small The phase diagrams for $\epsilon_N=4$. The thin red line is upper bounds obtained from $\epsilon_N=4$ fragmentation instabilities. These red lines covers larger area than the two fragmented cases.}} \label{fig4} \end{figure}

The AdS rotating black hole can be generally fragmented to more than two non-rotating black holes. The fragmentation depend on the number of non-rotating black holes. The sum of the black hole area will be smaller in the larger number of black holes. The number of non-rotating black holes is denoted as $\epsilon_N$, and the black hole mass is
\begin{eqnarray}
M=\epsilon_N\sqrt{m^2+\frac{J^2}{\epsilon_N^2 R^2}}\,.
\end{eqnarray}
The mass of final state is in terms of $\epsilon_N$\,,
\begin{eqnarray}
\mu'=\frac{2\mu}{(D-2)\epsilon_N\Xi^2}\sqrt{\left(1+\frac{D-4}{2}\Xi \right)^2-\frac{\Xi\,a^2}{r_h^2+a^2}}\,.
\end{eqnarray}
The dynamic Kerr bound moves upward in all of the phase diagrams. The fragmentation to two black hole gives the smallest area. The area increases as the number of fragmented black holes increases. Shown in Figure~\ref{fig4}\,, the shape of the digram is not changed, but the fragmentation occurs at larger spin parameter. The behaviors of the parameters are similar to all spacetime dimensions. Thus, the stable parameter ranges are minimum in the fragmentation of which the black hole is broken to two identical black holes without the mass loss.

\section{Conclusion and Discussion}
We investigated the instability of a rotating AdS black hole with fragmentation in large AdS radius limit. To avoid complexity, our investigation is valid in $\ell \gg R$ and sufficiently large distance between black holes to ignore gravitation in the final phase. In this approximation, the analysis is similar to that of asymptotically flat cases, and we can obtain overall behaviors of dynamic Kerr bounds in AdS spacetimes. Instability occurs before Kerr and BPS-like bounds for the general dimensions. It is given as a dynamic Kerr bound, which originated from the centrifugal force of the angular momentum. If the spin parameter $a$ is close to the AdS radius $\ell$\,, the centrifugal force increases the instability of the black hole, and then the black hole fragments. We constructed the process under energy and momentum conservation with an increase in entropy.

Black hole instability can be expected with an analytical approach. A black hole can be approximated to a small AdS Schwarzschild black hole to the $\mu\ll 1$ and $a\ll 1$\, case. The case is stable for fragmentation, because the ratio between initial and final state entropy is $\mathcal{A}_f/\mathcal{A}_i<1$\,. However, the large mass case $\mu\gg 1$ and $a\ll 1$ is $\mathcal{A}_f/\mathcal{A}_i>1$ and unstable. That means the centrifugal force is sufficiently large to break the black hole into pieces even if the spin parameter is small $a\ll 1$\,. Therefore, we can expect a mass upper bound in the approximation. In addition, a similar behavior appears in $\mu\ll 1$ and $a\sim \ell$\,. In the limit of $a \rightarrow \ell$\,, the final state black hole grows extremely large, so the final state entropy becomes sufficient to fragment the black hole. Thus, fragmentation instability exists, which suggests black hole parameter upper bounds called dynamic Kerr bounds. We solved eq.~(\ref{eq:entratio3}) with a numerical method and show detailed instability behaviors in redefined parameter space.

Instability is shown in detail in Figure~\ref{fig1}, \ref{fig2} under fragmentation. A 4-dimensional black hole exists stably under Kerr and dynamic Kerr bounds, and the cosmological constant increases the instability of a black hole, as shown in Figure~\ref{fig1}\,. Instability from the angular momentum becomes more dominant in more than 4 dimensions, which is shown in Figure~\ref{fig2} with the dynamic Kerr bound enclosing region $S_{FS}$ and $U_{S}$. The regions $S_{FS}$ and $U_{F}$ are stable under superradiance. Therefore, the black hole is stable in $S_{FS}$ region. It means that there are unstable region like $U_S$ in stable region under superradiance. Under fragmentation, dynamic Kerr bounds can limit the parameters of a black hole before BPS-like bounds, and a stable rotating AdS black hole exists within limited parameter ranges. Also, the centrifugal force becomes more important to instability than asymptotically flat case. If we increase the number of fragmented pieces and add the effect of the gravitational wave, the instability needs larger entropy and enlarges the dynamic Kerr area.

\vskip 0.20in
{\bf Acknowledgments}

This work was supported by the National Research Foundation of Korea(NRF) grant funded by the Korea government(MSIP)(No.~2014R1A2A1A010). BG was supported by Basic Science Research Program through the National Research Foundation of Korea(NRF) funded by the Ministry of Education, Science and Technology(NRF-2013R1A6A3A01020948).

\end{document}